\documentstyle[11pt,epsf]{article}
\begin{document}

\begin{center}

{\Large A Down to Earth Attempt at Determining the Low Energy Effective
Action of $N=2$ Supersymmetric Yang-Mills Theory \renewcommand
{\thefootnote}{\dagger}\footnote{contributed paper to LP99, August, SLAC, USA} }

\vspace{5mm}

{\large M. Chaichian${}^{a,b}$, 
W. F. Chen${}^{a,c}$ and C. Montonen${}^{b}$}

\vspace{5mm}

 {\it ${}^a$ High Energy Physics Division, Department of Physics, 
University of Helsinki\\
${}^b$ Helsinki Institute of Physics, 
P.O. Box 9 (Siltavuorenpenger 20 C)\\
FIN-00014 University of Helsinki, Finland \\
${}^c$ Winnipeg Institute for Theoretical Physics and
Department of Physics\\
University of Winnipeg, Winnipeg, Manitoba, Canada R3B 2E9}

\end{center}

\begin{abstract}
{\noindent We review a detailed investigation of the 
perturbative part of the low-energy effective 
action of $N=2$ supersymmetric Yang-Mills theory in a
conventional effective field theory approach. With the restriction 
that the effective action should contain at most two derivatives and 
not more than four-fermion couplings, the features of the low-energy
effective action obtained by Seiberg based on $U(1)_R$ anomaly and 
non-perturbative $\beta$-function arguments are shown to emerge.}
\end{abstract}
 
\vspace{1ex}

\section{Introduction}

The understanding to non-perturbative dynamics of supersymmetric gauge theory
has made rapid progress in recent years following the 
seminal contribution by Seiberg and Witten$\cite{sw}$, combining the ideas 
of holomorphicity$\cite{sei}$ and duality$\cite{mo}$.
The web of arguments leading to the explicit results consists of a skillful
combination of perturbative and nonperturbative arguments, formal
considerations and physical reasoning. It should be checked by explicit
computations, whenever possible, that no unexpected failure of these 
arguments occurs. 
In a recent work we have made an investigation in this
direction$\cite{ref4}$ and this paper is intended as a review.

The starting point in Seiberg and Witten's work is the low-energy
effective action of an $N=2$ supersymmetric Yang-Mills theory 
with the gauge group $SU(2)$ of the following form,
\begin{eqnarray}
\Gamma =\frac{1}{16\pi}\mbox{Im}{\int}d^4xd^2{\theta}d^2\widetilde{\theta}
\left[ \frac{1}{2}{\tau}{\Psi}^2
+\frac{i}{2\pi}{\Psi}^2\,\ln\frac{{\Psi}^2}{{\Lambda}^2}
+\sum_{n=1}^{\infty}A_n
\left(\frac{{\Lambda}^2}{{\Psi}^2}\right)^{2n}{\Psi}^2\right] \, ,
\label{eq1}
\end{eqnarray}
where $\displaystyle \tau=\frac{\theta}{2\pi}+\frac{4\pi i}{g^2}$
is the modular parameter and ${\Psi}$ the $N=2$ chiral superfield describing
the light degrees of freedom. The logarithmic term represents 
the one-loop perturbative result and was first obtained by Di Vecchia 
et al.${\cite{dmnp}}$ in a calculation where they coupled the gauge 
superfield to an $N=2$ matter supermultiplet and integrated out the 
latter. Subsequently, Seiberg$\cite{sei}$ used the anomalous transformation 
behaviour under $U(1)_R$ and holomorphicity to argue that the full 
low-energy effective action should take the form ({\ref{eq1}), where 
the infinite series arises from nonperturbative instanton contributions.
The Seiberg-Witten solution $\cite{sw}$ gives the explicit form of this 
part of $\Gamma$.

The form ({\ref{eq1}) has been confirmed by calculations in $N=1$ superspace
and in harmonic superspace, extending the result to nonleading terms in the
number of derivatives $\cite{dgr,pw,ov,ma,ke}$. Independent confirmation has
been obtained from $M$-theory $\cite{oo}$. 

Our intention is
to check the perturbative part of the effective 
action in Wess-Zumino gauge by a very down-to-earth calculation. In the Higgs 
phase of the theory, the $SU(2)$ gauge symmetry breaks down to $U(1)$, 
and the super-Higgs mechanism splits the supermultiplet into a massive 
one and a massless one. The effective action of the massless fields 
should be obtained by integrating out the heavy fields. In comparison
with other approaches, our method is quite conventional and is along
the lines of the standard definition of the low-energy effective theory. 

It should be emphasized that this conventional calculation is very
complicated. Even this modest programme we cannot carry out fully. 
What we have actually accomplished is the 
computation of the heavy fermion determinant. Reassuringly, we find that the form
(\ref{eq1}) is reproduced. Although no unexpected surprises were unearthed
by our calculation, we still hope that it has some pedagogical value in
showing explicitly how the effective action arises. 

The outline of this review is as follows. 
In section 2 we describe the model and exhibit the Higgs mechanism. Section 3
contains the computation of the heavy fermion determinant using the constant
field approximation. The detailed calculations of the fermion eigenvalues and
their degeneracies, which contain certain subtle points, are given 
in Appendix B. In section 4 we present a discussion of the results. 
In the pedagogical vein of this paper, we give in Appendix A the 
component form of the low-energy effective action (\ref{eq1}).

\section{Splitting of $N=2$ Supermultiplet }

The classical action of 
$N=2$ supersymmetric $SU(2)$ Yang-Mills theory is$\cite{dhd}$, 
\begin{eqnarray}
S&=& {\int}d^4 x\left[-\frac{1}{4}G_{\mu\nu}^aG^{\mu\nu a}
+D_{\mu}{\varphi}^{\dagger a}D^{\mu}{\varphi}^a
+i\overline{\psi}^a{\gamma}^{\mu}D_{\mu}^{ab}{\psi}^b\right.\nonumber\\[2mm]
&&+\left. \frac{ig}{\sqrt{2}}{\epsilon}^{abc}\overline{\psi}^c
[(1-{\gamma}_5){\varphi}^a
+(1+{\gamma}_5){\varphi}^{\dagger a}]{\psi}^b
+\frac{g^2}{2}{\epsilon}^{abc}{\epsilon}^{ade}
{\varphi}^b{\varphi}^{\dagger c}{\varphi}^{d}{\varphi}^{\dagger e}\right]\,,
\label{eq2} 
\end{eqnarray}
where
\begin{eqnarray}
G_{\mu\nu}^a&=&{\partial}_{\mu}K_{\nu}^a-{\partial}_{\nu}K_{\mu}^a
-g{\epsilon}^{abc}K_{\mu}^bK_{\nu}^c,~~
 D_{\mu}{\varphi}^a={\partial}_{\mu}{\varphi}^a
-g {\epsilon}^{abc}K_{\mu}^b{\varphi}^c,
\nonumber\\[2mm]
{\varphi}^a&=&\frac{1}{\sqrt{2}}(S^a+iP^a),~~ 
{\varphi}^{\dagger a}=\frac{1}{\sqrt{2}}(S^a-iP^a), ~~a=1,2,3\,.\nonumber
\end{eqnarray}
 The bosonic part of the action (\ref{eq2}) 
is just the Georgi-Glashow model
in the Bogomol'nyi-Prasad-Sommerfield (BPS) limit. 
In addition to the fermionic term and Yukawa interaction term, this action
 has the scalar potential
\begin{eqnarray}
V(\varphi)=-\frac{g^2}{2}{\epsilon}^{abc}{\epsilon}^{ade}
{\varphi}^b{\varphi}^{\dagger c}
{\varphi}^d{\varphi}^{\dagger e}
{\equiv}g^2\mbox{Tr}\left([\varphi,{\varphi}^{\dagger}]\right)^2\,.
\end{eqnarray}
The unbroken supersymmetry requires that in the ground state the
scalar potential must vanish, which leads to
\begin{eqnarray}
[\varphi,{\varphi}^{\dagger}]&=&0\,.
\label{eq:h2}
\end{eqnarray}
(\ref{eq:h2}) means that ${\varphi}^{\dagger}$ and ${\varphi}$ 
commute. Since the theory is gauge invariant, we can always choose$\cite{dhd}$  
\begin{eqnarray}
\langle S^a \rangle =v{\delta}^{a3},~~\langle P^a \rangle =0 \, ,
\label{eq:c1}
\end{eqnarray}
where $v$ is a real constant.   
For $v{\neq}0$ the theory is in the Higgs phase
and exhibits a spontaneous breaking of the gauge symmetry.
In a unitary gauge 
\begin{eqnarray}
S^{T}=\left(0,0, S+v \right)  \, .
\label{ug} 
\end{eqnarray}
The corresponding classical Lagrangian can be written as follows,
\begin{eqnarray}
{\cal L}={\cal L}_V+{\cal L}_S+{\cal L}_P+{\cal L}_F+{\cal L}_Y\, ,
\end{eqnarray}
where ${\cal L}_V$, ${\cal L}_S$, ${\cal L}_P$, ${\cal L}_F$ and ${\cal L}_Y$
denote respectively the vector field, the scalar field,
the scalar potential, 
the fermionic and the Yukawa interaction parts, 
\begin{eqnarray}
{\cal L}_V&=&  -\frac{1}{4}({\partial}^{\mu}A^{\nu}
-{\partial}^{\nu}A^{\mu}) ({\partial}_{\mu}A_{\nu}-{\partial}_{\nu}A_{\mu}) 
-\frac{1}{2}({\partial}_{\mu}W_{\nu}^+-{\partial}_{\nu}W_{\mu}^+) 
({\partial}^{\mu}W^{-\nu}-{\partial}^{\nu}W^{-\mu})\nonumber  \\[2mm]
&&-i g[({\partial}^{\mu}W_{\nu}^+W_{\mu}^-
-{\partial}^{\mu}W_{\nu}^-W_{\mu}^+)A^{\nu}+
({\partial}_{\nu}W_{\mu}^-W^{+\mu}
-{\partial}_{\nu}W^+_{\mu}W^{\mu -})A^{\nu}\nonumber\\[2mm]
&&+({\partial}^{\mu}A^{\nu}
-{\partial}^{\nu}A^{\mu})W_{\mu}^+W_{\nu}^-]
+g^2(-W_{\mu}^+W^{-\mu}A_{\nu}A^{\nu}+W_{\mu}^+W_{\nu}^-A^{\mu}A^{\nu})
\nonumber\\
&&+ \frac{g^2}{2}W^{+\mu}W^{-\nu}(W^+_{\mu}W^-_{\nu}-W^-_{\mu} W^+_{\nu});
\label{eq8}
\end{eqnarray}
\begin{eqnarray}
{\cal L}_S&=&\frac{1}{2}\partial_{\mu}P\partial^{\mu}P
+\frac{1}{2}\partial_{\mu}S\partial^{\mu}S+\partial_{\mu}P^+\partial^{\mu}P^-
+igA^{\mu}(\partial_{\mu}P^-P^+-\partial_{\mu}P^+P^-)\nonumber\\
&&+igP(\partial^{\mu}P^+W_{\mu}^--\partial^{\mu}P^-W_{\mu}^+)+
ig\partial^{\mu}P(W_{\mu}^+P^--W_{\mu}^-P^+)+g^2P^2W^{+\mu}W^-_{\mu}\nonumber\\
&&+g^2(S+v)^2W^{+\mu}W^-_{\mu}+g^2A^{\nu}A_{\nu}P^{+}P^--g^2(W_{\mu}^+P^--
W_{\mu}^-P^+)A^{\mu}P\nonumber\\
&&-\frac{g^2}{2}(W^{\mu +}P^--W^{\mu -}P^+)^2.
\end{eqnarray}
\begin{eqnarray}
{\cal L}_P=g^2(S+v)^2P^+P^-,
\end{eqnarray}
where the various quantities are defined as follows:
\begin{eqnarray}
&& W_{\mu}^+{\equiv}\frac{1}{\sqrt{2}}(K_{\mu}^{1}
-iK_{\mu}^{2})~,~
W_{\mu}^-{\equiv}\frac{1}{\sqrt{2}}(K_{\mu}^{1}
+iK_{\mu}^{2})~,~K_{\mu}^3 {\equiv}A_{\mu} \,;\nonumber\\
&&P^+ {\equiv}\frac{1}{\sqrt{2}}(P^1-iP^2)~,~P^-{\equiv}
 \frac{1}{\sqrt{2}}(P^1+iP^2)~,~P^3{\equiv}P.
\end{eqnarray}
The above Lagrangians  clearly show that $W_{\mu}^{\pm}$ and $P^{\pm}$
become massive with mass $m{\equiv}|gv|$ while $A_{\mu}$, $S$ and $P$
remain massless.

Up to some total derivative terms, the bosonic part of the Lagrangian 
can be rewritten in the following form:
\begin{eqnarray} 
{\cal L}_B&=&{\cal L}_V+{\cal L}_S+{\cal L}_P\nonumber\\
&=&-\frac{1}{4}F_{\mu\nu}F^{\mu\nu}+\frac{1}{2}\partial_{\mu}P\partial^{\mu}P
+\frac{1}{2}\partial_{\mu}S\partial^{\mu}S+\frac{1}{2}
W^{+\mu}\left[
{\eta}_{\mu\nu}D^{\dagger\alpha}D_{\alpha}-D_{\nu}^{\dagger}D_{\mu}
-igF_{\mu\nu}\right]W^{-\nu}\nonumber\\[2mm]
&&+\frac{1}{2}W^{-\mu}\left[
{\eta}_{\mu\nu}D^{\alpha}D^{\dagger}_{\alpha}
-D_{\nu}D_{\mu}^{\dagger}+igF_{\mu\nu}\right]W^{+\nu}
+g^2[P^2+(S+v)^2]W_{\mu}^+W^{\mu -}\nonumber\\
&&+\frac{1}{2}P^+(-\partial^{\mu}\partial_{\mu}
+2igA_{\mu}\partial^{\mu}+g^2A_{\mu}A^{\mu})P^-
+\frac{1}{2}P^-(-\partial^{\mu}\partial_{\mu}
-2igA_{\mu}\partial^{\mu}+g^2A_{\mu}A^{\mu})P^+
\nonumber\\
&&+\frac{1}{2}W_{\mu}^+(-igP\partial^{\mu}+ig\partial^{\mu}P-g^2A_{\mu}P)P^-
+\frac{1}{2}P^-(igP\partial^{\mu}+2ig\partial^{\mu}P-g^2A_{\mu}P)W_{\mu}^+
\nonumber\\
&&+\frac{1}{2}P^+(-2ig\partial^{\mu}P-igP\partial^{\mu}-g^2A^{\mu}P)W_{\mu}^-
+\frac{1}{2}W_{\mu}^-(-ig\partial^{\mu}P+igP\partial^{\mu}-g^2A^{\mu}P)P^+
\nonumber\\
&&+ \frac{g^2}{2}W^{+\mu}W^{-\nu}(W^+_{\mu}W^-_{\nu}-W^-_{\mu} W^+_{\nu})
-\frac{g^2}{2}(W^+_{\mu}P^--W^-_{\mu}P^+)^2\nonumber\\ 
&=&-\frac{1}{4}F_{\mu\nu}F^{\mu\nu}
+\partial_{\mu}{\phi}^{*}\partial^{\mu}{\phi}
+\frac{1}{2}W^{+\mu}{\Delta}_{\mu\nu}W^{-\nu}
+\frac{1}{2}W^{-\mu}{\Delta}^{\dagger}_{\mu\nu}W^{+\nu}
+\frac{1}{2}P^+\Delta P^-\nonumber\\
&&+\frac{1}{2}P^-\Delta^{\dagger}P^+
+\frac{1}{2}W^{+\mu}\Delta_{\mu}P^-
+\frac{1}{2}P^-\widetilde{\Delta}_{\mu}W^{+\mu}
+\frac{1}{2}P^+\widetilde{\Delta}_{\mu}^{\dagger}W^{-\mu}
+\frac{1}{2}W^{-\mu}\Delta_{\mu}^{\dagger}P^+\nonumber\\
&&+ \frac{g^2}{2}W^{+\mu}W^{-\nu}(W^+_{\mu}W^-_{\nu}-W^-_{\mu} W^+_{\nu})
-\frac{g^2}{2}(W^+_{\mu}P^--W^-_{\mu}P^+)^2 \, ,
\end{eqnarray} 
where 
\begin{eqnarray} 
{\Delta}_{\mu\nu}
&{\equiv}&{\eta}_{\mu\nu}D^{\dagger\alpha}D_{\alpha}-D_{\nu}^{\dagger}D_{\mu}
-igF_{\mu\nu}+g^2|\sqrt{2}\phi+v|^2{\eta}_{\mu\nu},\nonumber\\[2mm]
{\Delta}_{\mu\nu}^{\dagger}
&=&{\eta}_{\mu\nu}D^{\alpha}D_{\alpha}^{\dagger}
-D_{\nu}D_{\mu}^{\dagger}
+igF_{\mu\nu}+g^2|\sqrt{2}\phi+v|^2{\eta}_{\mu\nu};\nonumber\\[2mm]
\Delta_{\mu}&{\equiv}&-igP\partial^{\mu}+ig\partial^{\mu}P-g^2A_{\mu}P,~
\widetilde{\Delta}_{\mu}{\equiv}igP\partial^{\mu}
+2ig\partial^{\mu}P-g^2A_{\mu}P,
\nonumber\\
\Delta_{\mu}^{\dagger}&=&-ig\partial^{\mu}P+igP\partial^{\mu}-g^2A^{\mu}P,
~\widetilde{\Delta}_{\mu}^{\dagger}
=-2ig\partial^{\mu}P-igP\partial^{\mu}-g^2A^{\mu}P;\nonumber\\
\Delta &=& -\partial^{\mu}\partial_{\mu}
+2igA_{\mu}\partial^{\mu}+g^2A_{\mu}A^{\mu},~
\Delta^{\dagger}=-\partial^{\mu}\partial_{\mu}
-2igA_{\mu}\partial^{\mu}+g^2A_{\mu}A^{\mu}; \nonumber\\
D_{\mu}&=&\partial_{\mu}-igA_{\mu}, ~D_{\mu}^{\dagger}
=\partial_{\mu}+igA_{\mu}; ~\phi{\equiv}\frac{1}{\sqrt{2}}(S+iP).	
\end{eqnarray}

To explicitly show that the spinor fields split into massive and massless ones, 
we need some operations on ${\cal L}_F$ and ${\cal L}_Y$. The fermionic 
part is
\begin{eqnarray}
{\cal L}_F&=&i\overline{\psi}^1{\gamma}^{\mu}{\partial}_{\mu}{\psi}^1+
i\overline{\psi}^2{\gamma}^{\mu}{\partial}_{\mu}{\psi}^2
+i\overline{\psi}^3{\gamma}^{\mu}{\partial}_{\mu}{\psi}^3\nonumber\\[2mm]
&&+ \frac{g}{\sqrt{2}}\overline{\psi}^1(W_{\mu}^+-W_{\mu}^-)
{\gamma}^{\mu}{\psi}^3+ig \overline{\psi}^1A_{\mu}{\gamma}^{\mu}{\psi}^2
\nonumber\\[2mm]
&&+ig \overline{\psi}^1A_{\mu}{\gamma}^{\mu}{\psi}^2+
 \frac{ig}{\sqrt{2}}\overline{\psi}^2
{\gamma}^{\mu}(W_{\mu}^++W_{\mu}^-){\psi}^3
\nonumber\\[2mm]
&&-\frac{ig}{\sqrt{2}}\overline{\psi}^3{\gamma}^{\mu}
(W_{\mu}^++W_{\mu}^-){\psi}^2-
\frac{g}{\sqrt{2}}\overline{\psi}^3{\gamma}^{\mu}
(W_{\mu}^+-W_{\mu}^-){\psi}^1 \, .
\label{fp}
\end{eqnarray}
As for the Yukawa part, we first write it in terms of chiral spinors,
\begin{eqnarray}
{\cal L}_Y=i\sqrt{2}gf^{abc}\overline{\psi}_L^c{\varphi}^a{\psi}_R^b+i
\sqrt{2}gf^{abc}\overline{\psi}_R^c{\varphi}^{\dagger  a}{\psi}_L^b \, ,
\label{yu}
\end{eqnarray}
where ${\psi}_L=\displaystyle \frac{1}{2}(1-{\gamma}_5){\psi}$ and 
${\psi}_R=\displaystyle \frac{1}{2}(1+{\gamma}_5){\psi}$.
In the unitary gauge, Eq.(\ref{yu}) becomes
\begin{eqnarray}
{\cal L}_Y&=&ig(\sqrt{2}{\phi}+v)
(\overline{\psi}_L^2{\psi}_R^1-\overline{\psi}_L^1{\psi}_R^2)
+ig (\sqrt{2}{\phi}^*+v) (\overline{\psi}_R^2{\psi}_L^1
-\overline{\psi}_R^1{\psi}_L^2)\nonumber\\
&&+\frac{g}{\sqrt{2}}(P^++P^-)\left[(\overline{\psi}_R^3{\psi}_L^2
-\overline{\psi}_R^2{\psi}_L^3)-(\overline{\psi}_L^3{\psi}_R^2
-\overline{\psi}_L^2{\psi}_R^3)\right]\nonumber\\
&&+\frac{ig}{\sqrt{2}}(P^+-P^-)\left[(\overline{\psi}_R^1{\psi}_L^3
-\overline{\psi}_R^3{\psi}_L^1)-(\overline{\psi}_L^1{\psi}_R^3
-\overline{\psi}_L^3{\psi}_R^1)\right] \, .
\end{eqnarray}
With the combination
\begin{eqnarray}
{\Psi}_1{\equiv} \frac{1}{\sqrt{2}}({\psi}^1+i{\psi}^2)\,,\,
   {\Psi}_2{\equiv} \frac{1}{\sqrt{2}}({\psi}^1-i{\psi}^2)\, ,\,
{\Psi}{\equiv}{\psi}^3 \, , 
\end{eqnarray}
${\cal L}_F$ and ${\cal L}_Y$ can be formulated in these new fields, 
\begin{eqnarray}
{\cal L}_Y&=&
-g\overline{\Psi}_1\left[\frac{1}{\sqrt{2}}(1-\gamma_5)\phi
+\frac{1}{\sqrt{2}}(1+\gamma_5)\phi^*+v\right]{\Psi}_1\nonumber\\
&&+g\overline{\Psi}_2 \left[\frac{1}{\sqrt{2}}(1-\gamma_5)\phi
+\frac{1}{\sqrt{2}}(1+\gamma_5)\phi^*+v\right]{\Psi}_2
\nonumber\\
&&- igP^+\overline{\Psi}\gamma_5\Psi_1+igP^-\overline{\Psi}\gamma_5\Psi_2
-ig\overline{\Psi}_1 \gamma_5\Psi P^-+ig\overline{\Psi}_2\gamma_5\Psi P^+ \, ,
\\[2mm]
{\cal L}_F&=&i\overline{\Psi}_1{\gamma}^{\mu}{\partial}_{\mu}{\Psi}_1+
i\overline{\Psi}_2{\gamma}^{\mu}{\partial}_{\mu}{\Psi}_2
+i\overline{\Psi}{\gamma}^{\mu}{\partial}_{\mu}{\Psi}\nonumber\\[2mm]
&&+g\overline{\Psi}_1{\gamma}^{\mu}A_{\mu}{\Psi}_1-
g\overline{\Psi}_2{\gamma}^{\mu}A_{\mu}{\Psi}_2 \nonumber\\[2mm]
&&+g\overline{\Psi}_2{\gamma}^{\mu}W_{\mu}^+{\Psi}-
g\overline{\Psi}_1{\gamma}^{\mu}W_{\mu}^-{\Psi}\nonumber\\[2mm]
&&-g\overline{\Psi}{\gamma}^{\mu}W_{\mu}^+{\Psi}_1+
g\overline{\Psi}{\gamma}^{\mu}W_{\mu}^-{\Psi}_2\, .
\end{eqnarray}
So now the whole classical action is given by the Lagrangian
\begin{eqnarray}
{\cal L}&=&-\frac{1}{4}F_{\mu\nu}F^{\mu\nu}+\partial_{\mu}\phi^*
\partial^{\mu}\phi +i\overline{\Psi}{\gamma}^{\mu}{\partial}_{\mu}{\Psi}+
\frac{1}{2}W^{+\mu}{\Delta}_{\mu\nu}W^{-\nu}
+\frac{1}{2}W^{-\mu}{\Delta}^{\dagger}_{\mu\nu}W^{+\nu}\nonumber\\
&&+\frac{1}{2}P^+\Delta P^-+\frac{1}{2}P^-{\Delta}^{\dagger}P^+
+\frac{1}{2}W^{+\mu}\Delta_{\mu}P^-+\frac{1}{2}P^-
\widetilde{\Delta}_{\mu}W^{+\mu}
\nonumber\\
&&+\frac{1}{2}P^+\widetilde{\Delta}_{\mu}^{\dagger}W^{-\mu}
+\frac{1}{2}W^{-\mu}\Delta_{\mu}^{\dagger}P^+ 
+\overline{\Psi}_1\Delta_F{\Psi}_1
+\overline{\Psi}_2\widetilde{\Delta}_F{\Psi}_2
\nonumber \\[2mm]
&&- igP^+\overline{\Psi}\gamma_5\Psi_1+igP^-\overline{\Psi}\gamma_5\Psi_2
-ig\overline{\Psi}_1 \gamma_5\Psi P^-+ig\overline{\Psi}_2\gamma_5\Psi P^+ 
\nonumber\\[2mm]
&&+ g\overline{\Psi}_2{\gamma}^{\mu}W_{\mu}^+{\Psi}-
g\overline{\Psi}_1{\gamma}^{\mu}W_{\mu}^-{\Psi}
-g\overline{\Psi}{\gamma}^{\mu}W_{\mu}^+{\Psi}_1+
g\overline{\Psi}{\gamma}^{\mu}W_{\mu}^-{\Psi}_2\nonumber\\
&&+ \frac{g^2}{2}W^{+\mu}W^{-\nu}(W^+_{\mu}W^-_{\nu}-W^-_{\mu} W^+_{\nu})
-\frac{g^2}{2}(W^+_{\mu}P^--W^-_{\mu}P^+)^2 \, 
\label{eq20}
\end{eqnarray}
with
\begin{eqnarray}
\Delta_F&{\equiv}&i{\gamma}^{\mu}D_{\mu}
-\frac{g}{\sqrt{2}}(1-\gamma_5)\phi-\frac{g}{\sqrt{2}}(1+\gamma_5)\phi^*-gv,
\nonumber\\
\widetilde{\Delta}_F&{\equiv}&i{\gamma}^{\mu}D_{\mu}^{\dagger}
+\frac{g}{\sqrt{2}}(1-\gamma_5)\phi +\frac{g}{\sqrt{2}}(1+\gamma_5)\phi^*+gv.
\end{eqnarray}

\section{Low-energy Effective Action: Calculation of the Fermionic 
Determinant in Constant Field Approximation}

The standard definition of the low-energy effective action is given by 
\begin{eqnarray}
\mbox{exp}\left\{i\,{\Gamma}_{\rm eff}
[A_{\mu},{\phi},{\Psi},\overline{\Psi}]\right\}
{\equiv}{\int} {\cal D} W_{\mu}^+ {\cal D} W_{\mu}^-
{\cal D}\overline{\Psi}_1{\cal D}\overline{\Psi}_2 {\cal D}{\Psi}_1
{\cal D}{\Psi}_2 {\cal D}P^+{\cal D}P^- \mbox{exp}
\left[i\, \int \, d^4 x\, {\cal L}\right]\, .
\end{eqnarray}
At tree level
\begin{eqnarray}
{\Gamma}_{\rm eff}^{(0)}=S_{\rm tree}={\int}\,d^4x\left[ -\frac{1}{4}
F_{\mu\nu}F^{\mu\nu}
+{\partial}^{\mu}{\phi}^*{\partial}_{\mu}{\phi}+
i\overline{\Psi}{\gamma}^{\mu}{\partial}_{\mu}{\Psi}\right]\,.
\end{eqnarray}
At one-loop level, the integration over the heavy modes 
will lead to the determinants of the dynamical operators. 
In practical calculation we cannot evaluate
the determinant exactly. Here we shall
employ a technique called constant field 
approximation to compute the determinant, which
was invented by Schwinger${\cite{sch}}$ 
and later was used in
in \cite{dmnp} and \cite{ddds} to extract 
the anomaly term in $N=2$ supersymmetric Yang-Mills
theory and the one-loop effective action of the supersymmetric $CP^{N-1}$ model.
To apply this method
 we first rewrite the the quadratic part of the
 classical action (\ref{eq20}) as 
\begin{eqnarray}
S_{\rm quad}=S_{\rm tree}+{\int}d^4x\left(\Phi^{\dagger} M_{bb} \Phi+
\overline{\widetilde{\Psi}}M_{fb}\Phi +\Phi^{\dagger}M_{bf}\widetilde{\Psi}
+\overline{\widetilde{\Psi}}M_{ff}\widetilde{\Psi}\right),
\end{eqnarray}
where
\begin{eqnarray}
\Phi&{\equiv}&\left(\begin{array}{c} W^{-\mu}\\W^{+\mu}\\P^-\\P^+\end{array}\right),
~~\widetilde{\Psi}{\equiv}\left(\begin{array}{c}\Psi_1\\ \Psi_2\\
\Psi_1\\ \Psi_2\\  \end{array}\right);\nonumber\\
M_{bb}&{\equiv}&\frac{1}{2}
\left(\begin{array}{cccc}
{\Delta}_{\mu\nu} & 0 &\Delta_{\mu} & 0\\
0 & {\Delta}^{\dagger}_{\mu\nu} &0 &\Delta_{\mu}^{\dagger} \\
\widetilde{\Delta}_{\nu}^{\dagger} & 0 &\Delta &0 \\
0 & \widetilde{\Delta}_{\nu} & 0 & \Delta^{\dagger}
\end{array}\right);\nonumber\\
M_{fb}&{\equiv}&\frac{1}{2} \left(\begin{array}{cccc}
-g\gamma_{\mu}\Psi & 0 & -ig\gamma_5 \Psi & 0 \\
0 & g\gamma_{\mu}\Psi & 0 & ig\gamma_5 \Psi  \\
-g\gamma_{\mu}\Psi & 0 &  -ig\gamma_5 \Psi & 0\\
0 & g\gamma_{\mu}\Psi & 0 &  ig\gamma_5 \Psi\\
 \end{array}\right); \nonumber\\
M_{bf}&{\equiv}&\frac{1}{2} \left(\begin{array}{cccc}
-g\overline{\Psi}\gamma_{\mu} & 0 & -g\overline{\Psi}\gamma_{\mu} & 0  \\
0 &g\overline{\Psi}\gamma_{\mu} & 0 & g\overline{\Psi}\gamma_{\mu} \\
-ig\overline{\Psi} \gamma_5 & 0 &  -ig\overline{\Psi} \gamma_5 & 0\\
0 & ig\overline{\Psi} \gamma_5 & 0 & ig\overline{\Psi} \gamma_5
\end{array}\right);\nonumber\\
M_{ff}&{\equiv}&\frac{1}{2} \left(\begin{array}{cccc}
\Delta_F & 0 & 0 & 0\\
0 & \widetilde{\Delta}_F & 0 & 0 \\
0 & 0 & \Delta_F & 0 \\
0 & 0 & 0 &\widetilde{\Delta}_F
 \end{array}\right).
\label{eq27x}
\end{eqnarray}
Using the standard formulas
\begin{eqnarray} 
I&=&\int {\cal D}b^{\dagger}{\cal D}b{\cal D}\overline{f}{\cal D}f\,
\mbox{exp}\left[\int (dx)
\left(b^{\dagger}M_{bb}b+\overline{f}M_{fb}b+b^{\dagger}M_{bf}f
+\overline{f}M_{ff}f\right)\right] \nonumber\\[2mm]
&=&\int{\cal D}b^{\dagger}{\cal D}b{\cal D}\overline{f}{\cal D}f\,
 \mbox{exp}\left\{\int (dx) \left[b^{\dagger}(M_{bb}-M_{bf}M_{ff}^{-1}M_{fb})b
\right.\right.\nonumber\\
&&+\left.\left.(\overline{f}
+b^{\dagger}M_{bf}M_{ff}^{-1})M_{ff}(M_{ff}^{-1}M_{fb}b+f)
\right]\right\}
\nonumber\\[2mm]
&=&\det M_{ff}\det{}^{-1}(M_{bb}-M_{bf}M_{ff}^{-1}M_{fb}); \nonumber\\
\det M&=&\mbox{exp}\,\mbox{Tr}\,\ln M,
\end{eqnarray}
$b$ and $f$ representing the general 
bosonic and fermionic fields, respectively,
we get
\begin{eqnarray}
Z[A, \phi,\Psi,\overline{\Psi}]
&=&\mbox{exp}\left\{i\,{\Gamma}_{\rm eff}
[A_{\mu},{\phi},{\Psi},\overline{\Psi}]\right\}
{\equiv}{\int} {\cal D} W_{\mu}^+ {\cal D} W_{\mu}^-
{\cal D}\overline{\Psi}_1{\cal D}\overline{\Psi}_2 {\cal D}{\Psi}_1
{\cal D}{\Psi}_2 \mbox{exp}\left[ iS \right] \nonumber\\
&=& \mbox{exp}\left[iS_{\rm tree}\right]\det M_{ff}\det{}^{-1}(M_{bb}
-M_{bf}M_{ff}^{-1}M_{fb})\nonumber\\
&=&\mbox{exp}\left[iS_{\rm tree}+\mbox{Tr}\ln M_{ff}-
\mbox{Tr}\ln(M_{bb}-M_{bf}M_{ff}^{-1}M_{fb})\right];\nonumber\\
{\Gamma}_{\rm eff}&=&S_{\rm tree}-i\left[\mbox{Tr}\ln M_{ff}-
\mbox{Tr}\ln(M_{bb}-M_{bf}M_{ff}^{-1}M_{fb})\right].
\label{eq27} 
\end{eqnarray}

The following task is to evaluate the above determinants. 
Let us first consider the fermionic 
part. Since $M_{ff}$  has the form of a reducible matrix,
\begin{eqnarray} 
\det M_{ff}=\frac{1}{16}(\det\Delta_F)^2(\det\widetilde{\Delta}_F)^2
=\frac{1}{16}\mbox{exp}[2(\mbox{Tr}\ln\Delta_F
+\mbox{Tr}\ln\det\widetilde{\Delta}_F)].
\end{eqnarray}
Now we switch on the constant field approximation to work out the eigenvalues
and eigenvectors of the above operators and further evaluate the determinant.
We choose only the third components of the electric
and magnetic fields to be the constants different from zero,
\begin{eqnarray} 
-E_3=F^{03}{\neq}0, ~~B_3=F^{12}{\neq}0,
\end{eqnarray}
and $\phi$ the non-vanishing constant field.
Consequently, the potential becomes
\begin{eqnarray}                
A^1=-F^{12}x_2, ~~A^3=-F^{30}x_0, ~~A^0=A^2=0.
\end{eqnarray}
To get the eigenvalues of the operators, it is necessary
to rotate into Euclidean space,
\begin{eqnarray}
&& x^4=x_4=-ix^0, ~~\partial_0=\frac{\partial}{\partial x^0}
=i\frac{\partial}{\partial x^4},\nonumber\\
&& f^{34}=f_{34}=iF^{30}, ~~f^{12}=f_{12}=F^{12}.
\end{eqnarray}
Let us first consider $\det \Delta_F$.
The eigenvalue equation for $\Delta_F$ is
\begin{eqnarray}
\Delta_F\psi(x)=\left[i\gamma^{\mu}D_{\mu}-\frac{g}{\sqrt{2}}(1-\gamma_5)\phi
-\frac{g}{\sqrt{2}}(1+\gamma_5)\phi^* -gv\right]\psi(x) 
=\omega \psi_1,
\label{eq32}
\end{eqnarray}
where $\psi$ is a four-component spinor wave function.
In order to get normalizable eigenstates, we consider the system in a
box of finite size $L$ in the $x_1$ and $x_3$ directions with periodic
boundary conditions, so the eigenvector should be of the following form,
\begin{eqnarray}
\psi(x)=\frac{1}{L}e^{ip_1{\cdot}x_1}e^{ip_3{\cdot}x_3}{\chi}(x_2,x_4), 
\nonumber\\
p_1=\frac{2\pi l}{L}, ~~p_3=\frac{2\pi k}{L},~~k,l=\mbox{integers}.
\end{eqnarray}
To find the eigenvalues and eigenvectors, 
we write the operators and the wave function in two-component forms,
\begin{eqnarray}
\Delta_F=\left(\begin{array}{cc} -g(\sqrt{2}\phi^*+v){\bf 1} & \Delta^-\\
\Delta^+ & -g(\sqrt{2}\phi +v){\bf 1} \end{array}\right),~~~
\chi =\left(\begin{array}{c}\chi_{1}\\ \chi_{2}\end{array}\right),
\end{eqnarray}
where ${\bf 1}$ is the $2{\times}2$ identity matrix and
\begin{eqnarray}
\Delta^{\pm}=\partial_4{\pm}i\left[\sigma_1(\partial_1+igf_{12}x_2)
+\sigma_2\partial_2+\sigma_3(\partial_3+igf_{34}x_4)\right].
\end{eqnarray}
The eigenvalue equation (\ref{eq32}) 
is thus reduced to the following set of equations,
\begin{eqnarray}
-g(\sqrt{2}\phi^*+v){\chi}_{1}+ \Delta^-{\chi}_{2}
&=&\omega {\chi}_{1},\nonumber\\
\Delta^+{\chi}_{1}-g(\sqrt{2}\phi +v){\chi}_{2}&=&\omega {\chi}_{2},
\label{eq54}
\end{eqnarray}
and now
\begin{eqnarray}
\Delta^{\pm}=\partial_4{\mp}[\sigma_1(p_1+gf_{12}x_2)
-i\sigma_2\partial_2+\sigma_3(p_3+gf_{34}x_4)].
\end{eqnarray}
A detailed calculation and discussion of the eigenvalues 
are collected in Appendix B.
We obtain two series of eigenvalues,
\begin{eqnarray}
\omega_{\pm}(m,n)
=-g\left[\frac{(\phi +\phi^*)}{\sqrt{2}} +v\right]
{\pm}\sqrt{\frac{1}{2}g^2(\phi -\phi^*)^2-2mgf_{12}-2ngf_{34}},
\end{eqnarray} 
where for $m{\geq}1$, $n{\geq}1$ both eigenvalues are doubly degenerate,
while $\omega_{\pm}(m.0)$ and $\omega_{\pm}(0,n)$ are nondegenerate, 
and for $m=n=0$, there exists only a nondegenerate eigenvalue 
$\omega_-(0,0)$.

In a similar way we can solve the eigenvalue equation 
\begin{eqnarray}
\widetilde{\Delta}_F\widetilde{\psi}=\left[i\gamma^{\mu}D_{\mu}^{\dagger}
+\frac{g}{\sqrt{2}}(1-\gamma_5)\phi 
+\frac{g}{\sqrt{2}}(1+\gamma_5)\phi^*+gv \right]\widetilde{\psi}
=\widetilde{\omega}\widetilde{\psi},
\end{eqnarray}
 and obtain the eigenvalues,
\begin{eqnarray}
\widetilde{\omega}_{\pm}(m,n)
=g\left[\frac{(\phi +\phi^*)}{\sqrt{2}} +v\right]
{\pm}\sqrt{\frac{g^2}{2}(\phi -\phi^* )^2-2m gf_{12}-2n gf_{34}}.
\end{eqnarray}
The degeneracies of $\widetilde{\omega}_{\pm}(m,n)$, 
$\widetilde{\omega}_{\pm}(m,0)$ and $\widetilde{\omega}_{\pm}(0,n)$ with
$m{\geq}1$, $n{\geq}1$ are the same as those of the $\omega$s. There
 still only exists a nondegenerate eigenvalue $\omega_-(0,0)$.    

With the above eigenvalues
$\mbox{Tr}\ln\Delta_F$ and $\mbox{Tr}\ln\widetilde{\Delta}_F$ can be computed
straightforwardly, 
\begin{eqnarray}
\mbox{Tr}\ln\Delta_F=\ln\det\Delta_F
=\ln\left[\Pi\omega_{\pm (lk)}(m,n)\right]^r
=\sum_{l,k=-\infty}^{+\infty}\sum_{m,n=0}^{\infty} r
\ln\omega_{\pm (lk)}(m,n),
\end{eqnarray}
where $r$ is the degeneracy of $\omega_{\pm}(m,n)$.
Due to the relation $x_2=2\pi l/(gf_{12} L)$ and $x_4=2\pi k/(gf_{34} L)$,
the summation over the momenta $k$ and $l$ is actually equivalent to
an integration over $x_2$ and $x_4$. Since the fields are constants,
this integration will yield only a Euclidean 
space volume factor, which tends to infinity
in the continuous limit ($L{\rightarrow}\infty$), 
\begin{eqnarray}
\sum_{l,k}=\frac{L^2}{4\pi^2}g^2f_{12}f_{34}{\int}dx_2dx_4
=\frac{V}{4\pi^2}g^2f_{12}f_{34},
\end{eqnarray}
while the Lagrangian will be well defined.
Consider the degeneracy of each eigenvalue, we have
\begin{eqnarray}
\mbox{Tr}\ln\Delta_F&=&\frac{V}{4\pi^2}g^2f_{12}f_{34}
\left\{\left[\ln\omega_-(0,0)+\sum_{m=1}^{\infty}\ln\omega_+(m,0)
+\sum_{n=1}^{\infty}\ln\omega_+(0,n)\right.\right.\nonumber\\
&+&\left.\left.2\sum_{m,n=1}^{\infty}\ln\omega_+(m,n)
\right] +\left[\sum_{m=1}^{\infty}\ln\omega_-(m,0)+
\sum_{n=1}^{\infty}\ln\omega_-(0,n)
+2\sum_{m,n=1}^{\infty}\ln\omega_-(m,n)
\right]\right\}\nonumber\\
&=&\frac{V}{4\pi^2}ge^2f_{12}f_{34}\left\{\ln\omega_-(0,0)
+\sum_{m=1}^{\infty}\ln[\omega_+(m,0)\omega_-(m,0)]
\right.\nonumber\\
&&+\left.\sum_{n=1}^{\infty}\ln[\omega_+(0,n)\omega_-(0,n)]
+2\sum_{m,n=1}^{\infty}\ln\left[\omega_+(m,n)\omega_-(m,n)\right]
\right\} \nonumber\\ 
&=&\frac{V}{4\pi^2}g^2f_{12}f_{34}
\left\{\ln[-g(\sqrt{2}\phi +v)]+
\sum_{m=1}^{\infty}
\ln [g^2(\sqrt{2}\phi^*+v)(\sqrt{2}\phi+v)+2mgf_{12}]\right.\nonumber\\
&&+\sum_{n=1}^{\infty}\ln [g^2(\sqrt{2}\phi^*+v)(\sqrt{2}\phi+v)
+2ngf_{34}]\nonumber\\
&&+\left.2\sum_{m,n=1}^{\infty}
\ln [g^2(\sqrt{2}\phi^*+v)(\sqrt{2}\phi+v)+2mgf_{12}+2ngf_{34}]\right\}.
\end{eqnarray}
Similarly, we get
\begin{eqnarray}
\mbox{Tr}\ln\widetilde{\Delta}_F
&=&\frac{V}{4\pi^2}g^2f_{12}f_{34}\left\{\ln\left[g(\sqrt{2}\phi^*+v) \right]
+\sum_{m=1}^{\infty}
\ln [g^2(\sqrt{2}\phi^*+v)(\sqrt{2}\phi+v)+2mgf_{12}]\right.\nonumber\\
&&+\sum_{n=1}^{\infty}\ln [g^2(\sqrt{2}\phi^*+v)(\sqrt{2}\phi+v)+2ngf_{34}]
\nonumber\\
&&+\left.2\sum_{m,n=1}^{\infty}\ln [g^2(\sqrt{2}\phi^*+v)(\sqrt{2}\phi+v)
+2mgf_{12}+2ngf_{34}]\right\}.
\end{eqnarray}
Thus we finally obtain
\begin{eqnarray}
\mbox{Tr}\ln\Delta_F +\mbox{Tr}\ln\widetilde{\Delta}_F
&=&\frac{V}{4\pi^2}g^2f_{12}f_{34}
\left\{\ln\left[\frac{\sqrt{2}\phi^*+v }{\sqrt{2}\phi +v }
\right]\right.\nonumber\\
&&+2\sum_{m=1}^{\infty} \ln [g^2(\sqrt{2}\phi^*+v)(\sqrt{2}\phi+v)
+2mgf_{12}]\nonumber\\
&&+2\sum_{n=1}^{\infty}\ln [g^2(\sqrt{2}\phi^*+v)(\sqrt{2}\phi+v)
+2ngf_{34}]\nonumber\\
&&+ \left.4\sum_{m,n=1}^{\infty}\ln [g^2(\sqrt{2}\phi^*+v)(\sqrt{2}\phi+v)
+2mgf_{12}+2ngf_{34}]  \right\}.
\end{eqnarray}
Making using of the proper-time regularization,
\begin{eqnarray}
\ln \alpha=-\int^{\infty}_{{1}/{\Lambda^2}}\frac{ds}{s}e^{-\alpha s}
\end{eqnarray}
with $\Lambda^2$ being the cut-off to regularize the infinite sum,
we have
\begin{eqnarray}
&&\mbox{Tr}\ln\Delta_F +\mbox{Tr}\ln\widetilde{\Delta}_F
=\frac{V}{4\pi^2}g^2f_{12}f_{34}\left\{
\ln\left[\frac{\sqrt{2}\phi^*+v}{\sqrt{2}\phi+v) }\right]
-2\int^{\infty}_{{1}/{\Lambda^2}}
\frac{ds}{s}e^{-g^2(\sqrt{2}\phi^*+v)(\sqrt{2}\phi+v) s}\right.\nonumber\\
&&{\times}\left.\left[\sum_{m=1}^{\infty}e^{-2mgf_{12}s}
+\sum_{n=1}^{\infty}e^{-2ngf_{34}s}
+2\sum_{m,n=1}^{\infty}e^{-(2mgf_{12}+2ngf_{34})s}\right]\right\}
\nonumber\\
&=&\frac{V}{4\pi^2}g^2f_{12}f_{34}
\left\{\ln\left[\frac{\sqrt{2}\phi^*+v }{\sqrt{2}\phi+v }\right]
-2\int^{\infty}_{{1}/{\Lambda^2}}
\frac{ds}{s}e^{-g^2(\sqrt{2}\phi^*+v)(\sqrt{2}\phi+v) s}
\left[\frac{e^{-gf_{34}s}}{\sinh (gf_{34}s)}\right.\right.\nonumber\\
&&+\frac{e^{-gf_{12}s}}{\sinh (gf_{12}s)}
+\left.\left.\frac{e^{-(gf_{12}+gf_{34})s}}
{\sinh (gf_{12}s)\,\sinh (gf_{34}s)}\right]\right\}
\nonumber\\
&=&\frac{V}{4\pi^2}g^2f_{12}f_{34}
\left[\ln\left(\frac{\sqrt{2}\phi^*+v}{\sqrt{2}\phi+v }\right)
-\int^{\infty}_{{1}/{\Lambda^2}}\frac{ds}{s}
e^{-g^2(\sqrt{2}\phi^*+v)(\sqrt{2}\phi+v)s}\coth(gf_{12}s)
\coth(gf_{34}s)\right]\nonumber\\
&=&\frac{V}{4\pi^2}g^2f_{12}f_{34}
\left[\ln\left(\frac{\sqrt{2}\phi^*+v}{\sqrt{2}\phi+v}\right)\right.\nonumber\\
&&-\left.\int^{\infty}_{{1}/{\Lambda^2}}\frac{ds}{s}
e^{-g^2(\sqrt{2}\phi^*+v)(\sqrt{2}\phi+v) s}
\frac{\cosh[g(f_{12}+f_{34})s]+\cosh[g(f_{12}-f_{34})s]}
{\cosh[g(f_{12}+f_{34})s]-\cosh[g(f_{12}-f_{34})s]}\right]\, ,
\label{eq50n}
\end{eqnarray}
where we have used
\begin{eqnarray}
\sum_{m=1}^{\infty}e^{-2mt}=\frac{e^{-t}}{2\sinh t},~~~
\cosh (x{\pm}y)=\cosh x\cosh y{\pm}\sinh x\sinh y.
\end{eqnarray}
Rotating back to Minkowski space and denoting ${\bf X}{\equiv}{\bf H}+i{\bf E}$,
we write (\ref{eq50n}) as
\begin{eqnarray}
&&\mbox{Tr}\ln\Delta_F +\mbox{Tr}\ln\widetilde{\Delta}_F
=\frac{V}{4\pi^2}g^2iE_z H_z
\left[\ln\left(\frac{\sqrt{2}\phi^*+v }{\sqrt{2}\phi+v}\right)\right.
\nonumber\\
&-&\left.\int^{\infty}_{{1}/{\Lambda^2}}
\frac{ds}{s}e^{-g^2(\sqrt{2}\phi^*+v)(\sqrt{2}\phi+v) s}
\frac{ \cosh[g(H_z+iE_z)s]+\cosh[g(H_z-iE_z)s]}
{\cosh[g(H_z+iE_z)s]-\cosh[g(H_z-iE_z)s]}\right]\nonumber\\
&{\equiv}&\frac{V}{4\pi^2}g^2i{\bf E}{\cdot}{\bf H}
\left[\ln\left(\frac{\sqrt{2}\phi+v }{\sqrt{2}\phi^*+v }\right)
\nonumber\right.\\
&&-\left.\int^{\infty}_{{1}/{\Lambda^2}}\frac{ds}{s}
e^{-g^2(\sqrt{2}\phi^*+v)(\sqrt{2}\phi+v) s}
\frac{ \cosh[g{\bf X}s]+\cosh (g{\bf X}^*s)}
{\cosh (g{\bf X} s)-\cosh (g{\bf X}^*s)}\right],
\label{eq84}
\end{eqnarray}
To extract the divergence, we must analyze the small-$s$ behaviour of
the integrand of (\ref{eq84}) by
using the identities
\begin{eqnarray} 
i{\bf E}{\cdot}{\bf H}=\frac{1}{4}({\bf X}^2-{\bf X}^{*2})
=\frac{1}{4}F_{\mu\nu}\widetilde{F}^{\mu\nu}, ~~
{\bf H}^2-{\bf E}^2=\frac{1}{2}({\bf X}^2+{\bf X}^{*2})
=\frac{1}{2}F_{\mu\nu}F^{\mu\nu},
\end{eqnarray}
and the series expansion near $s\sim 0$
\begin{eqnarray} 
\frac{ \cosh[g{\bf X}s]+\cosh[g {\bf X}^*s]}
{\cosh[g{\bf X} s]-\cosh[g{\bf X}^*s]}
&=&\frac{1}{({\bf X}^2-{\bf X}^{*2})}\left[\frac{4}{g^2s^2}+
\frac{2}{3}({\bf X}^2+{\bf X}^{*2})+{\cal O}(s^2)\right].
\label{eq84x}
\end{eqnarray}
It can be easily seen from (\ref{eq84x}) that the integral in 
(\ref{eq84}) has a quadratic divergence and a logarithmic one. Thus
the divergence term can be extracted by writing (\ref{eq84}) 
as the form,
\begin{eqnarray}
&&\mbox{Tr}\ln\Delta_F +\mbox{Tr}\ln\widetilde{\Delta}_F
=\frac{V}{4\pi^2}\left\{\frac{1}{4}g^2F_{\mu\nu}\widetilde{F}^{\mu\nu}
 \ln\left[\frac{\sqrt{2}\phi^*+v}{\sqrt{2}\phi +v}\right]\right.\nonumber\\
&-&\int^{\infty}_{{1}/{\Lambda^2}}ds
\left(\frac{1}{s^3}+\frac{1}{6}\frac{1}{s}g^2F_{\mu\nu}F^{\mu\nu}\right)
e^{-g^2(\sqrt{2}\phi^*+v)(\sqrt{2}\phi+v)s}\nonumber\\
&-&\int^{\infty}_0\frac{ds}{s^3}
e^{-g^2(\sqrt{2}\phi^*+v)(\sqrt{2}\phi+v) s}\left[\frac{1}{4}g^2s^2
F_{\mu\nu}\widetilde{F}^{\mu\nu}\frac{ \cosh (g{\bf X}s)+\cosh (g{\bf X}^*s)}
{\cosh (g{\bf X} s)-\cosh (g{\bf X}^*s)}\right.\nonumber\\
&-&\left.\left.\frac{1}{s^3}
-\frac{1}{6}\frac{1}{s}g^2F_{\mu\nu}F^{\mu\nu}\right]\right\}.
\label{eq86}
\end{eqnarray}
The second term (\ref{eq86}) is the UV divergent term, 
so the cut-off $1/\Lambda^2$ is preserved to regularize the 
integral, while the last term is a finite term
and hence the cut-off has been removed.

Now we turn to the bosonic determinant. From (\ref{eq27}) we have  
\begin{footnotesize}
\begin{eqnarray}
&&M_{ff}^{-1}=2\left(\begin{array}{cccc} 1/{\Delta}_F & 0 & 0 & 0\\
0 & 1/\widetilde{\Delta}_F & 0 & 0 \\
0 & 0 & 1/{\Delta}_F & 0 \\
0 & 0 & 0 & 1/\widetilde{\Delta}_F \end{array}\right), ~~
M_{bb}-M_{bf}M_{ff}^{-1}M_{fb}=\nonumber\\ &&  \frac{1}{2}
\left(\begin{array}{cccc} \Delta_{\mu\nu}
-2g^2\overline{\psi}{\gamma}_{\mu}\frac{1}{\widetilde{\Delta}_F}{\gamma}_{\nu}\psi & 0 &
\Delta_{\mu}-2ig^2\overline{\psi}{\gamma}_{\mu}\frac{1}{\Delta_F}{\gamma}_5\psi & 0\\
0 & {\Delta}^{\dagger}_{\mu\nu}
-2g^2\overline{\psi}{\gamma}_{\mu}\frac{1}{\widetilde{\Delta}_F}{\gamma}_{\nu}\psi & 0
& {\Delta}^{\dagger}_{\mu}
-2ig^2\overline{\psi}{\gamma}_{\mu}\frac{1}{\widetilde{\Delta}_F}{\gamma}_5\psi\\
\widetilde{\Delta}_{\nu}^{\dagger}-2ig^2\overline{\psi}{\gamma}_5
\frac{1}{\Delta_F}{\gamma}_{\nu}\psi & 0 & \Delta+2g^2\overline{\psi}{\gamma}_5
\frac{1}{\Delta_F}{\gamma}_5\psi & 0\\
0 &  \widetilde{\Delta}_{\nu}-2ig^2\overline{\psi}{\gamma}_5
\frac{1}{\Delta_F}{\gamma}_{\nu}\psi & 0 & \Delta^{\dagger}
+2g^2\overline{\psi}{\gamma}_5
\frac{1}{\Delta_F}{\gamma}_5\psi
\end{array}\right).\nonumber\\
\end{eqnarray}
\end{footnotesize}
In constant field approximation, $\overline{\psi}$ and $\psi$ can be regarded
Grassman numbers, so we can expand the bosonic determinant only 
to the quartic terms in  $\overline{\psi}$ and $\psi$.
Now the key problem is how to find the eigenvalues and eigenstates of
the operator matrix $M_{bb}-M_{bf}M_{ff}^{-1}M_{fb}$. If they could be worked
out, then with the eigenvalues and eigenvectors of fermionic 
operator, we can use the technique 
developed in \cite{dmnp} to evaluate this determinant. Unfortunately,
it seems to us that in the constant field approximation
it is very to find the eigenvalues and eigenstates of such 
a horrible operator matrix. This difficulty is waiting to be overcome.

Despite the fact that the bosonic part cannot be evaluated, we can see from 
(\ref{eq27}) and (\ref{eq86}) that the effective
Lagrangian associated with the fermionic part has already shown
the features of the perturbative part of the low-energy effective action.
First, we believe that the quadratic divergence of Eq.(\ref{eq86})
will be canceled owing to the nonrenormalization theorem. 
Second, for the logarithmic divergence of Eq.(\ref{eq86}), with
\begin{eqnarray}
\int^{\infty}_{{1}/{\Lambda^2}}\frac{ds}{s}
e^{-g^2(\sqrt{2}\phi +v)(\sqrt{2}\phi^*+v) s}{\sim}
-\ln\left[\frac{g^2(\sqrt{2}\phi +v)(\sqrt{2}\phi^*+v) }{\Lambda^2}\right],
\end{eqnarray}
Eq.(\ref{eq86}) shows that the Wilson effective action has one term
proportional to
\begin{eqnarray}
F_{\mu\nu}F^{\mu\nu}\ln\left[
\frac{g^2(\sqrt{2}\phi^*+v)(\sqrt{2}\phi+v)}{\Lambda^2}\right].
\end{eqnarray}
Comparing with the component field form given by (\ref{eq:last}), 
we can conclude that the complete calculation should give the 
form (\ref{eq1}) of the low-energy effective action. One can even guess 
this from the requirement of
supersymmetry since the constant field approximation and the 
proper-time regularization preserve the supersymmetry explicitly.  
Further, there is a finite term proportional 
to $F\widetilde{F}\ln\left[(\sqrt{2}\phi +v)/(\sqrt{2}\phi^*+v)\right]$ in
(\ref{eq86}). As pointed out in \cite{dmnp}, this is the 
reflection of the axial $U(1)_R$ anomaly in the effective action. 
This anomaly term had played a crucial role
in the nonperturbative analysis\cite{sei}.
 
\section{Summary}

In summary, we have tried to calculate the perturbative part of
the low-energy effective action of 
$N=2$ supersymmetric Yang-Mills
theory based on a standard effective field theory technique.
It is well known that the Seiberg-Witten effective action is the
cornerstone for all those new developments in $N=2$ supersymmetric
gauge theory, and that this effective action has been obtained
in an indirect way. Therefore, it is worthwhile to
try to compute this effective action using a straightforward
integration of the heavy degrees of freedom.
 Unfortunately, we have encountered an insurmountable difficulty 
in evaluating the bosonic operator adopting the constant field
approximation. This prevents us from getting the  complete result and
giving a thorough comparison with the form of (\ref{eq1}). However, 
the calculation of the fermionic determinant has indeed shown the 
basic features of the low-energy effective action. This gives 
a partial verification of the abstract symmetry analysis in extracting 
the low-energy effective action. The complete calculation presents 
an interesting problem for further investigation.

\vspace{8mm}

\noindent{\bf Acknowledgments:}
 We acknowledge the financial support by the Academy of Finland 
under the project No. 37599 and 44129. W.F.C is partially supported  
by the Natural Sciences and Engineering Research Council of Canada. 
\newpage

\appendix

\section{Low-energy Effective Action in Component Field Form}

\setcounter{equation}{0}
\renewcommand{\theequation}{A.\arabic{equation}}

 To compare our result with that obtained  from non-perturbative analysis,
in this appendix we write the perturbative part of the Seiberg-Witten 
low-energy effective action (\ref{eq1}) in the form of component
fields.  
First,  (\ref{eq1}) can be expressed in $N=1$ superfield
\begin{eqnarray}
{\Gamma}= \frac{1}{16\pi}\mbox{Im} {\int}d^4 x\left[{\int}
d^2{\theta}{\cal F}''(\Phi)W^{\alpha}W_{\alpha}+{\int}d^2{\theta}
d^2\overline{\theta}{\Phi}^{\dagger}{\cal F}'(\Phi)\right]\, ,
\end{eqnarray}
where ${\Phi}$ is the $N=1$ chiral superfield
\begin{eqnarray} 
 {\Phi}={\phi}(x)+i{\theta}{\sigma}^{\mu}
\overline{\theta}{\partial}_{\mu}{\phi}-\frac{1}{4}{\theta}^2
\overline{\theta}^2{\partial}^2{\phi}+\sqrt{2}{\theta}{\psi}
-\frac{i}{\sqrt{2}}{\theta}^2{\partial}_{\mu}{\psi}
{\sigma}^{\mu}\overline{\theta}+{\theta}^2F(x) \, ,
\end{eqnarray}
and 
\begin{eqnarray} 
{\cal F}(\Phi)=\frac{1}{2} {\tau}{\Phi}^2+
\frac{i}{2\pi}{\Phi}^2\ln\frac{{\Phi}^2}{{\Lambda}^2} \,,~~~\tau =
 \frac{\theta}{2\pi}+\frac{4{\pi}i}{g^2}\,. 
\label{pre1}
\end{eqnarray}
In Wess-Zumino gauge, the Abelian vector superfield and the 
corresponding superfield strength are, respectively, 
\begin{eqnarray}
V&=&-{\theta}{\sigma}^{\mu}\overline{\theta}A_{\mu}+i{\theta}^2
(\overline{\theta}\overline{\lambda})-i{\overline{\theta}}^2(\theta\lambda)+
\frac{1}{2}{\theta}^2{\overline{\theta}}^2D\, ;\nonumber\\
W_{\alpha}&=&-i{\lambda}_{\alpha}(y)+{\theta}_{\alpha}D-
i{\sigma}_{\alpha}^{\mu\nu\beta}{\theta}^{\beta}F_{\mu\nu}(y)
+{\theta}^2{\sigma}^{\mu\beta}_{\alpha}{\partial}_{\mu}
\overline{\lambda}_{\beta}(y)\,,
\end{eqnarray}
where $y^{\mu}
=x^{\mu}+i{\theta}{\sigma}^{\mu}\overline{\theta}$, ${\sigma}^{\mu\nu}
=\frac{1}{4}({\sigma}^{\mu}\overline{\sigma}^{\nu}-
{\sigma}^{\nu}\overline{\sigma}^{\mu})$ and $F_{\mu\nu}={\partial}_{\mu}A_{\nu}
-{\partial}_{\nu}A_{\mu}$.
Using the expansion
\begin{eqnarray}
{\cal F}(\Phi)&=& {\cal F}(\phi)+{\cal F}'(\phi) \left[
i{\theta}{\sigma}^{\mu}
\overline{\theta}{\partial}_{\mu}{\phi}-\frac{1}{4}{\theta}^2
\overline{\theta}^2{\partial}^2{\phi}+\sqrt{2}{\theta}{\psi}
-\frac{i}{\sqrt{2}}{\theta}^2{\partial}_{\mu}{\psi}
{\sigma}^{\mu}\overline{\theta}+{\theta}^2F(x)\right]\nonumber\\[2mm]  
&&+ \frac{1}{2}{\cal F}''(\phi) \left[
i{\theta}{\sigma}^{\mu}
\overline{\theta}{\partial}_{\mu}{\phi}-\frac{1}{4}{\theta}^2
\overline{\theta}^2{\partial}^2{\phi}+\sqrt{2}{\theta}{\psi}
-\frac{i}{\sqrt{2}}{\theta}^2{\partial}_{\mu}{\psi}
{\sigma}^{\mu}\overline{\theta}
+{\theta}^2F(x)\right]\nonumber\\[2mm] 
&&{\times}\left[i{\theta}{\sigma}^{\mu}
\overline{\theta}{\partial}_{\mu}{\phi}-\frac{1}{4}{\theta}^2
\overline{\theta}^2{\partial}^2{\phi}+\sqrt{2}{\theta}{\psi}
-\frac{i}{\sqrt{2}}{\theta}^2{\partial}_{\mu}{\psi}
{\sigma}^{\mu}\overline{\theta}+{\theta}^2F(x)\right],
\end{eqnarray}
and the similar expansion for ${\cal F}'(\Phi)$,  we obtain
\begin{eqnarray}
{\Gamma}&=& \frac{1}{16 \pi}\mbox{Im} {\int}d^4x\,\left[
-{\cal F}''(\phi){\phi}{\partial}^2{\phi}-{\cal F}^{(3)}(\phi){\phi}
{\partial}^{\mu}{\phi}{\partial}_{\mu}{\phi}
+2 {\cal F}''(\phi){\partial}_{\mu}{\phi}{\partial}^{\mu}{\phi}\right.
\nonumber\\[2mm]
&&-{\partial}^2{\phi}{\cal F}'(\phi)  
+2i{\cal F}''(\phi){\partial}_{\mu}{\psi}{\sigma}^{\mu}\overline{\psi}
-2i{\cal F}''(\phi){\psi}{\sigma}^{\mu}{\partial}_{\mu}\overline{\psi}
+2i{\cal F}^{(3)}(\phi){\psi}{\sigma}^{\mu}\overline{\psi}{\partial}_{\mu}{\phi}
\nonumber\\[2mm]
&&-2 {\cal F}^{(3)}(\phi)F^{\dagger}{\psi}{\psi}+4F^{\dagger}F{\cal F}''(\phi)
+4i {\cal F}''(\phi){\lambda} {\sigma}^{\mu}{\partial}_{\mu}\overline{\lambda}-2
{\cal F}''(\phi)D^2\nonumber\\[2mm]
&&+4 {\cal F}''(\phi) (-F^{\mu\nu}F_{\mu\nu}
+iF_{\mu\nu}\widetilde{F}^{\mu\nu})-2 \sqrt{2}i {\cal F}^{(3)}(\phi)
{\psi}{\lambda}D+2{\cal F}^{(3)}(\phi)({\lambda}{\lambda})F\nonumber\\[2mm]
&&-\left.{\cal F}^{(4)}(\phi)({\psi}{\psi})({\lambda}{\lambda})\right] \, .
\label{sw}
\end{eqnarray}
Using (\ref{pre1}) 
\begin{eqnarray}
{\cal F}'(\phi)=\left({\tau}+\frac{i}{\pi}\right){\phi}+
\frac{i}{\pi}{\phi}\ln\frac{\phi^2}{{\Lambda}^2} ~&,&~
{\cal F}''(\phi)={\tau}+\frac{3i}{\pi}+
\frac{i}{\pi}\ln\frac{\phi^2}{{\Lambda}^2}\, ,
\nonumber\\[2mm]
{\cal F}^{(3)}(\phi)=\frac{2i}{\pi}\,\frac{1}{\phi} ~&,&~
{\cal F}^{(4)}(\phi)=-\frac{2i}{\pi}\,\frac{1}{{\phi}^2}\,,
\end{eqnarray}
and rescaling the field $X{\longrightarrow}gX$, $X=(A,{\phi},
{\lambda},{\psi})$, we write (\ref{sw}) as 
\begin{eqnarray}
{\Gamma}&=& \int d^4x\left\{[-8{\pi}{\phi}{\partial}^2{\phi}+
8{\pi}{\partial}_{\mu}{\phi}{\partial}^{\mu}{\phi}+8{\pi}\,i
{\partial}_{\mu}{\psi}\overline{\sigma}^{\mu}\overline{\psi}-8{\pi}\,i{\psi}
\overline{\sigma}^{\mu}{\partial}_{\mu}\overline{\psi}+16{\pi}\,i\lambda
{\sigma}^{\mu}{\partial}_{\mu}\overline{\lambda}\right.\nonumber\\[2mm]
&&-4{\pi}F_{\mu\nu}F^{\mu\nu}]
+\frac{g^2}{\pi}\,[-4{\phi}{\partial}^2{\phi}
+4{\partial}_{\mu}{\phi}{\partial}^{\mu}{\phi}+6\,i
{\partial}_{\mu}{\psi}\overline{\sigma}^{\mu}\overline{\psi}-6{\pi}\,i{\psi}
\overline{\sigma}^{\mu}{\partial}_{\mu}\overline{\psi}\nonumber\\[2mm]
&&+4\,i {\psi}\overline{\sigma}^{\mu}\overline{\psi}{\partial}_{\mu}{\phi}
\frac{1}{\phi}+12\,i\lambda
{\sigma}^{\mu}{\partial}_{\mu}\overline{\lambda}-3\,F_{\mu\nu}F^{\mu\nu}]
+\frac{g^2}{\pi}\,
\ln\frac{\phi^2}{\Lambda^2}
[-2{\phi}{\partial}^2{\phi}
+2{\partial}_{\mu}{\phi}{\partial}^{\mu}{\phi}\nonumber\\[2mm]
&&+2\,i{\partial}_{\mu}{\psi}\overline{\sigma}^{\mu}\overline{\psi}
-2\,i{\psi}\overline{\sigma}^{\mu}{\partial}_{\mu}\overline{\psi}
+4 \,i\lambda{\sigma}^{\mu}{\partial}_{\mu}\overline{\lambda}
-F_{\mu\nu}F^{\mu\nu}]+\frac{g^2}{8{\pi}^2}\frac{1}{{\phi}^2}
(\psi\psi)(\lambda\lambda)\nonumber\\[2mm]
&&-\frac{1}{2}
\left(1+\frac{3g^2}{4{\pi}^2}+\frac{g^2}{8{\pi}^2}
\ln\frac{{\phi}^2}{{\Lambda}^2}\right)
 -\frac{\sqrt{2}g^2}{4{\pi}^2}
\frac{i(\psi\lambda)D}{\phi}-\frac{g^2}{4{\pi}^2}\frac{F^{\dagger}
(\psi\psi)}{\phi}
\nonumber\\[2mm]
&&+F^{\dagger}F\left(1+\frac{3g^2}{4{\pi}^2}
+\frac{g^2}{4{\pi}^2}\ln\frac{\phi ^2}{
\lambda ^2}\right)+\frac{g^2}{4\pi ^2}\frac{(\lambda\lambda)F}{\phi},
\label{eqa8}
\end{eqnarray}
where the vacuum angle ${\theta}$ is set to zero. Eliminating the auxiliary
fields $F$, $F^{\dagger}$ and $D$ with the equations of motion 
derived from (\ref{eqa8}),
\begin{eqnarray}
F \left(1+\frac{3g^2}{4{\pi}^2}+\frac{g^2}{8{\pi}^2}
\ln\frac{{\phi}^2}{{\Lambda}^2}\right)-\frac{g^2}{4{\pi}^2}
\frac{\psi\psi}{\phi}&=&0, \nonumber\\[2mm]
F^{\dagger}\left(1+\frac{3g^2}{4{\pi}^2}+\frac{g^2}{8{\pi}^2}
\ln\frac{{\phi}^2}{{\Lambda}^2}\right)+\frac{g^2}{4{\pi}^2}
\frac{\lambda\lambda}{\phi}&=&0,\nonumber\\[2mm]
D\left(1+\frac{3g^2}{4{\pi}^2}+\frac{g^2}{8{\pi}^2}
\ln\frac{{\phi}^2}{{\Lambda}^2}\right)+i\frac{\sqrt{2}g^2}{4{\pi}^2}
\frac{\psi\lambda}{\phi}&=&0,
\end{eqnarray}
and performing the algebraic manipulations
\begin{eqnarray}
{\int}d^4x\,\ln\frac{g^2{\phi}^2}{{\Lambda}^2}
{\phi}{\partial}^2{\phi}
&=& -{\int}d^4x\,\left[2+ \ln\frac{g^2{\phi}^2}{{\Lambda}^2}\right]
{\partial}_{\mu}{\phi}{\partial}^{\mu}{\phi},\\
{\int}d^4x\,{\psi}\overline{\sigma}^{\mu}\overline{\psi}\frac{{\partial}_{\mu}
\phi}{\phi}
&=& -\frac{1}{2}{\int}d^4x\,
\ln\frac{g^2{\phi}^2}{{\Lambda}^2}\,[
{\partial}_{\mu}{\psi}\overline{\sigma}^{\mu}\overline{\psi}
+{\psi}\overline{\sigma}^{\mu}{\partial}_{\mu}\overline{\psi}]\, ,
\end{eqnarray}
we get
\begin{eqnarray}
{\Gamma}&=&  \int d^4x\left\{\left[{\partial}_{\mu}{\phi}
{\partial}^{\mu}{\phi}+{\lambda}
{\sigma}^{\mu}{\partial}_{\mu}\overline{\lambda}+i\overline{\psi}\overline{\sigma}^{\mu}
{\partial}_{\mu}{\psi}
-\frac{1}{4}F_{\mu\nu}F^{\mu\nu}\right]\right.\nonumber\\[2mm]            
&&+ \frac{3g^2}{4 {\pi}^2}\left[
{\partial}_{\mu}{\phi}{\phi}+\overline{\lambda}
{\sigma}^{\mu}{\partial}_{\mu}\overline{\lambda}+i\overline{\psi}\overline{\sigma}^{\mu}
{\partial}_{\mu}{\psi}-\frac{1}{4}F_{\mu\nu}F^{\mu\nu}\right]
\nonumber\\[2mm]
&&+\frac{g^2}{4{\pi}^2}
\ln\frac{g^2\phi^2}{{\Lambda}^2}\left[ 
{\partial}_{\mu}{\phi}{\partial}^{\mu}{\phi}+i\overline{\lambda}
{\sigma}^{\mu}{\partial}_{\mu}\overline{\lambda}+i\overline{\psi}\overline{\sigma}^{\mu}
{\partial}_{\mu}{\psi}-\frac{1}{4}
F_{\mu\nu}F^{\mu\nu}\right] 
+\left.\frac{g^2}{8{\pi}^2}\frac{(\lambda\lambda)(\psi\psi)}{\phi^2}
\right\} \,. 
\label{swe}
\end{eqnarray}         
Considering the four-component spinor field form
$$ {\Psi}=\left(\begin{array}{l}
\psi\\ 
\overline{\lambda} \end{array} \right)\,, ~\overline{\Psi}=(\lambda , \overline{\psi})\,, ~
{\gamma}^{\mu}=\left(\begin{array}{ll}
0 & {\sigma}^{\mu}\\
\overline{\sigma}^{\mu} & 0 \end{array} \right)\, , $$
and especially using the fact that for a $N=2$ Abelian supermultiplet, 
$\Psi$ should be a Majorana spinor: $\psi=\lambda$ and
 $\overline{\psi}=\overline{\lambda}$
 and $(\lambda\lambda)(\psi\psi)=(\psi\psi)^2
=1/4(\overline{\Psi}{\Psi})^2$, we finally write
 (\ref{swe}) as the following form 
\begin{eqnarray}
{\Gamma}= \int d^4x\left\{\left[1+\frac{3g^2}{4{\pi}^2}
+\frac{g^2}{4{\pi}^2}\ln\frac{\phi^2}{{\Lambda}^2}\right]\,\left[
{\partial}_{\mu}{\phi}
{\partial}^{\mu}{\phi}+
i\overline{\Psi}\overline{\gamma}^{\mu}
{\partial}_{\mu}{\Psi}-\frac{1}{4}F_{\mu\nu}F^{\mu\nu}\right]
+\frac{g^4}{32{\pi}^2}\frac{(\overline{\Psi}{\Psi})^2}{g^2{\phi}^2}\right\}.
\label{eq:last}
\end{eqnarray}
(\ref{eq:last}) is the perturbative part of the low-energy effective 
action in Wess-Zumino gauge given by Seiberg$\cite{sei}$.

\section{Eigenvalues of the Fermionic Operator}

\setcounter{equation}{0}
\renewcommand{\theequation}{B.\arabic{equation}}

In this appendix we present a detailed calculation on the eigenvalues
of fermionic operator $\Delta_F$. First Eq.(\ref{eq54}) implies 
\begin{eqnarray}
\Delta^-\Delta^+{\chi}_{1}(x_2,x_4)
&=&\left[{\omega}+g(\sqrt{2}\phi^*+v)\right]
\left[{\omega}+g(\sqrt{2}\phi +v) \right]{\chi}_{1}(x_2,x_4),
\label{eq56} \nonumber\\
\Delta^+\Delta^-{\chi}_{2}(x_2,x_4)
&=&\left[{\omega}+g(\sqrt{2}\phi^*+v)\right]
\left[{\omega}+g(\sqrt{2}\phi +v) \right] {\chi}_{2}(x_2,x_4)
\label{eq57}
\end{eqnarray}
with
\begin{eqnarray}
\Delta^+\Delta^-
&=&-H_{12}-H_{34}+g\sigma_3(f_{12}+f_{34}),
\nonumber\\
\Delta^-\Delta^+&=&-H_{12}-H_{34}+g\sigma_3(f_{12}-f_{34}),
\end{eqnarray}
where $H_{12}$ and $H_{34}$ are the Hamiltonian operators of
two independent harmonic oscillators,
\begin{eqnarray}
H_{12}&=&-\frac{\partial^2}{\partial x^2_2}
+g^2f_{12}^2(x_2+\frac{p_1}{gf_{12}})^2=-\frac{\partial^2}{\partial \xi^2_2}
+\Omega_{12}^2\xi^2_2, ~\xi_2{\equiv}x_2+\frac{p_1}{gf_{12}},
~\Omega_{12}{\equiv}|gf_{12}|;\nonumber\\
H_{34}&=&-\frac{\partial^2}{\partial x^2_4}
+g^2f_{34}^2(x_4+\frac{p_3}{gf_{34}})^2=-\frac{\partial^2}{\partial \xi^2_4}
+\Omega_{34}^2\xi^2_4, ~\xi_4=x_4+\frac{p_3}{gf_{34}},
~\Omega_{34}=|gf_{34}|.
\end{eqnarray}
Eq.(\ref{eq57}) means that the eigenvalue and the eigenvector of
$\Delta_F$ must be that of $\Delta^+\Delta^-$ and $\Delta^-\Delta^+$, while
the reverse may be not true. In the following we make use of the eigenvalue
and the eigenvector of $\Delta^+\Delta^-$ and $\Delta^-\Delta^+$ to find
the ones of $\Delta_F$. As the usual operator method dealing with the harmonic
oscillator, defining the destruction and creation operators
\begin{eqnarray}
a_2&=&\frac{1}{\sqrt{2}}
\left(\sqrt{\Omega_{12}}\xi_2+\frac{1}{\sqrt{\Omega_{12}}}
\frac{\partial}{\partial\xi_2}\right), ~~a_2^{\dagger}
=\frac{1}{\sqrt{2}}\left(\sqrt{\Omega_{12}}\xi_2-\frac{1}{\sqrt{\Omega_{12}}}
\frac{\partial}{\partial\xi_2}\right),~~[a_2,a_2^{\dagger}]=1;\nonumber\\
a_4&=&\frac{1}{\sqrt{2}}
\left(\sqrt{\Omega_{34}}\xi_4+\frac{1}{\sqrt{\Omega_{34}}}
\frac{\partial}{\partial\xi_4}\right), ~~a_4^{\dagger}
=\frac{1}{\sqrt{2}}\left(\sqrt{\Omega_{34}}\xi_4+\frac{1}{\sqrt{\Omega_{34}}}
\frac{\partial}{\partial\xi_4}\right),~~[a_4,a_4^{\dagger}]=1,\nonumber\\
\end{eqnarray}
we obtain the Hamiltonian operators and their eigenstates 
in Fock space,
\begin{eqnarray}
H_{12}&=&\Omega_{12}(2a_2a_2^{\dagger}+1), ~~|n_{12}{\rangle}=
\frac{1}{\sqrt{n_{12}!}}(a_2^{\dagger})^{n_{12}}|0_{12}{\rangle},\nonumber\\ 
a_2|0_{12}{\rangle}&=&0,~~
H_{12}|n_{12}{\rangle}=\Omega_{12}(2n_{12}+1)|n_{12}{\rangle};\nonumber\\
H_{34}&=&\Omega_{34}(2a_4a_4^{\dagger}+1), ~~|n_{34}{\rangle}=
\frac{1}{\sqrt{n_{34}!}}(a_4^{\dagger})^{n_{34}}|0_{34}{\rangle},\nonumber\\ 
a_4|0_{34}{\rangle}&=&0,~~
H_{34}|n_{34}{\rangle}=\Omega_{34}(2n_{34}+1)|n_{34}{\rangle}.
\end{eqnarray}
The operators $\Delta^+$ and $ \Delta^-$
can be rewritten in terms of the destruction and creation operators,
\begin{eqnarray}
\Delta^+&=&i\sigma_2\frac{\partial}{\partial \xi_2}-gf_{12}\xi_2\sigma_1
+\frac{\partial}{\partial \xi_4}-gf_{34}\xi_4\sigma_3\nonumber\\
&=&\left(\begin{array}{cc}
\sqrt{\frac{\Omega_{34}}{2}}(a_4-a_4^{\dagger})
-\frac{gf_{34}}{\sqrt{2\Omega_{34}}}(a_4+a_4^{\dagger}) &
\sqrt{\frac{\Omega_{12}}{2}}(a_2-a_2^{\dagger})
-\frac{gf_{12}}{\sqrt{2\Omega_{12}}}(a_2+a_2^{\dagger}) \\
-\sqrt{\frac{\Omega_{12}}{2}}(a_2-a_2^{\dagger})
-\frac{gf_{12}}{\sqrt{2\Omega_{34}}}(a_2+a_2^{\dagger}) &
\sqrt{\frac{\Omega_{34}}{2}}(a_4-a_4^{\dagger})
-\frac{gf_{34}}{\sqrt{2\Omega_{34}}}(a_4+a_4^{\dagger}) \end{array}\right);
\nonumber\\
\Delta^-&=&-i\sigma_2\frac{\partial}{\partial \xi_2}+gf_{12}\xi_2\sigma_1
+\frac{\partial}{\partial \xi_4}+gf_{34}\xi_4\sigma_3\nonumber\\
&=&\left(\begin{array}{cc}
\sqrt{\frac{\Omega_{34}}{2}}(a_4-a_4^{\dagger})
+\frac{gf_{34}}{\sqrt{2\Omega_{34}}}(a_4+a_4^{\dagger}) &
-\sqrt{\frac{\Omega_{12}}{2}}(a_2-a_2^{\dagger})
+\frac{gf_{12}}{\sqrt{2\Omega_{12}}}(a_2+a_2^{\dagger}) \\
\sqrt{\frac{\Omega_{12}}{2}}(a_2-a_2^{\dagger})
+\frac{gf_{12}}{\sqrt{2\Omega_{12}}}(a_2+a_2^{\dagger}) &
\sqrt{\frac{\Omega_{34}}{2}}(a_4-a_4^{\dagger})
-\frac{gf_{34}}{\sqrt{2\Omega_{34}}}(a_4+a_4^{\dagger}) \end{array}\right).
\end{eqnarray}
There are four different cases that should be considered:
\begin{itemize}
\item[1.] ~$gf_{12}>0$, $gf_{34}>0$; $\Omega_{12}=gf_{12}$, 
$\Omega_{34}=gf_{34}$;
\begin{eqnarray}
\Delta^+ &=&\left(\begin{array}{cc}
-\sqrt{2\Omega_{34}}a_4^{\dagger} & -\sqrt{2\Omega_{12}}a_2^{\dagger}\\
-\sqrt{2\Omega_{12}}a_2 & \sqrt{2\Omega_{34}}a_4\end{array}\right), ~~
\Delta^- =\left(\begin{array}{cc}
\sqrt{2\Omega_{34}}a_4 & \sqrt{2\Omega_{12}}a_2^{\dagger}\\
\sqrt{2\Omega_{12}}a_2 & -\sqrt{2\Omega_{34}}a_4^{\dagger} \end{array}\right);
\end{eqnarray}
\item[2.] ~$gf_{12}>0$, $gf_{34}<0$; $\Omega_{12}=gf_{12}$, 
$\Omega_{34}=-gf_{34}$;
\begin{eqnarray}
\Delta^+ &=&\left(\begin{array}{cc}
\sqrt{2\Omega_{34}}a_4 & -\sqrt{2\Omega_{12}}a_2^{\dagger}\\
-\sqrt{2\Omega_{12}}a_2 & -\sqrt{2\Omega_{34}}a_4^{\dagger}
\end{array}\right), ~~
\Delta^- =\left(\begin{array}{cc}
-\sqrt{2\Omega_{34}}a_4^{\dagger} & \sqrt{2\Omega_{12}}a_2^{\dagger}\\
\sqrt{2\Omega_{12}}a_2 & -\sqrt{2\Omega_{34}}a_4 \end{array}\right);
\end{eqnarray}
\item[3.] ~$gf_{12}<0$, $gf_{34}>0$; $\Omega_{12}=-gf_{12}$, 
$\Omega_{34}=gf_{34}$;
\begin{eqnarray}
\Delta^+ &=&\left(\begin{array}{cc}
-\sqrt{2\Omega_{34}}a_4^{\dagger} & \sqrt{2\Omega_{12}}a_2\\
\sqrt{2\Omega_{12}}a_2^{\dagger} & -\sqrt{2\Omega_{34}}a_4
\end{array}\right), ~~
\Delta^- =\left(\begin{array}{cc}
\sqrt{2\Omega_{34}}a_4 & -\sqrt{2\Omega_{12}}a_2\\
-\sqrt{2\Omega_{12}}a_2^{\dagger} & -\sqrt{2\Omega_{34}}a_4^{\dagger} 
\end{array}\right);
\end{eqnarray}
\item[4.] ~$gf_{12}<0$, $gf_{34}>0$; $\Omega_{12}=-gf_{12}$, 
$\Omega_{34}=gf_{34}$;
\begin{eqnarray}
\Delta^+ &=&\left(\begin{array}{cc}
\sqrt{2\Omega_{34}}a_4 & \sqrt{2\Omega_{12}}a_2 \\
\sqrt{2\Omega_{12}}a_2^{\dagger} & -\sqrt{2\Omega_{34}}a_4^{\dagger}
\end{array}\right), ~~
\Delta^- =\left(\begin{array}{cc}
-\sqrt{2\Omega_{34}}a_4^{\dagger} & -\sqrt{2\Omega_{12}}a_2\\
-\sqrt{2\Omega_{12}}a_2^{\dagger} & \sqrt{2\Omega_{34}}a_4^{\dagger} 
\end{array}\right).
\end{eqnarray}
\end{itemize}
Now we look for the eigenvalues of $\Delta_F$ with aid of the eigenvalues of 
$\Delta^+\Delta^-$ and $\Delta^-\Delta^+$. From Eq.(\ref{eq57}) we have
\begin{eqnarray}
\Delta^-\Delta^+|{\chi}_{1}{\rangle}
&=&\left[{\omega}+g(\sqrt{2}\phi^*+v)\right]
\left[{\omega}+g(\sqrt{2}\phi +v) \right]|{\chi}_{1}{\rangle},
\nonumber\\
\Delta^+\Delta^-| {\chi}_{2}{\rangle}
&=&\left[{\omega}+g(\sqrt{2}\phi^*+v)\right]
\left[{\omega}+g(\sqrt{2}\phi +v) \right] |{\chi}_{2}{\rangle}.
\end{eqnarray}
The eigenstates ${\chi}_{1}{\rangle}$ and $|{\chi}_{2}{\rangle}$ should 
be the following form,
\begin{eqnarray}
|{\chi}_{i}{\rangle}{\sim}\left(\begin{array}{l}|k,l{\rangle}\\ 
|m,n{\rangle}\end{array}\right),~~i=1, 2,
\end{eqnarray}
where $|k,l{\rangle}{\equiv}|k{\rangle}|l{\rangle}$,
 $|m,n{\rangle}{\equiv}|m{\rangle}|n{\rangle}$, $k,m$ are the quantum numbers
of the harmonic oscillator $H_{12}$ and $l,n$ are those of $H_{34}$. 

We first consider the case {\bf 1}, since that
\begin{eqnarray}
\Delta^-\Delta^+\left(\begin{array}{l}|k,l{\rangle}\\ 
|m,n{\rangle}\end{array}\right)&=&
\left(\begin{array}{c}\left[
-2k \Omega_{12}-2(l+1) \Omega_{34}\right]|k,l{\rangle} \\[1mm] 
\left[-2(m+1) \Omega_{12}-2n \Omega_{34}\right]
|m,n{\rangle} \end{array}\right),\nonumber\\
\Delta^+\Delta^-\left(\begin{array}{l}|k,l{\rangle}\\ 
|m,n{\rangle}\end{array}\right)&=&
\left(\begin{array}{c} \left(-2k\Omega_{12}-2l\Omega_{34}\right)
|k,l{\rangle}\\[1mm] 
\left[-2(m+1)\Omega_{12}-2(n+1)\Omega_{34}\right]
|m,n{\rangle}\end{array}\right),
\end{eqnarray}
the common eigenstate of $\Delta^-\Delta^+$ and $\Delta^+\Delta^-$
with eigenvalue $-2m\Omega_{12}-2n\Omega_{34}$, $m,n{\geq}1$ is
\begin{eqnarray}
\left(\begin{array}{c}|\chi_1{\rangle}\\ |\chi_2{\rangle}\end{array}\right)
=\left(\begin{array}{c}\left(\begin{array}{c}
\alpha |m,n-1{\rangle}\\ \beta|m-1,n{\rangle}\end{array}\right)\\
\left(\begin{array}{c}
\gamma |m,n{\rangle}\\ \delta|m-1,n-1{\rangle}\end{array}\right)\end{array}
\right),
\end{eqnarray}
where $\alpha$, $\beta$, $\gamma$ and $\delta$ are normalization parameters.
With this eigenstate, we rewrite the eigenvalue equation (\ref{eq54}) in 
Fock space,
\begin{eqnarray}
&&\left(\begin{array}{cc} -g(\sqrt{2}\phi^*+v){\bf 1} & \Delta^-\\
\Delta^+ & -g(\sqrt{2}\phi +v){\bf 1} \end{array}\right)
\left(\begin{array}{c}\left(\begin{array}{c}
\alpha |m,n-1{\rangle}\\ \beta|m-1,n{\rangle}\end{array}\right)\\
\left(\begin{array}{c}
\gamma |m,n{\rangle}\\ \delta|m-1,n-1{\rangle}\end{array}\right)\end{array}
\right)\nonumber\\
&&=\omega \left(\begin{array}{c}\left(\begin{array}{c}
\alpha |m,n-1{\rangle}\\ \beta|m-1,n{\rangle}\end{array}\right)\\
\left(\begin{array}{c}
\gamma |m,n{\rangle}\\ \delta|m-1,n-1{\rangle}\end{array}\right)\end{array}
\right)\nonumber\\
&&=\left(\begin{array}{c}\left[-g(\sqrt{2}\phi^*+v)\alpha +\sqrt{2n\Omega_{34}}
\gamma+\sqrt{2m\Omega_{12}}\delta \right]|m,n-1{\rangle}\\[1mm]
\left[-g(\sqrt{2}\phi^*+v)\beta +\sqrt{2m\Omega_{12}}
\gamma-\sqrt{2n\Omega_{34}}\delta \right]|m-1,n{\rangle}\\[1mm]
\left[-\sqrt{2n\Omega_{34}}\alpha-\sqrt{2m\Omega_{12}}\beta 
-g(\sqrt{2}\phi+v)\gamma \right]|m,n{\rangle}\\[1mm]
\left[-\sqrt{2m\Omega_{12}}\alpha+\sqrt{2n\Omega_{34}}\beta 
-g(\sqrt{2}\phi+v)\delta \right]|m-1,n-1{\rangle}\end{array}\right).
\label{eqb15}
\end{eqnarray}
Eq.(\ref{eqb15}) means that searching for the operator eigenvalue can be
 changed into an ordinary matrix eigenvalue problem,
\begin{eqnarray}
\left(\begin{array}{cccc} -g(\sqrt{2}\phi^*+v) & 0 & \sqrt{2n\Omega_{34}}
 & \sqrt{2m\Omega_{12}} \\[1mm]
 0 &-g(\sqrt{2}\phi^*+v)  & \sqrt{2m\Omega_{12}}
 & -\sqrt{2n\Omega_{34}} \\[1mm]
-\sqrt{2n\Omega_{34}} & -\sqrt{2m\Omega_{12}} & -g(\sqrt{2}\phi +v) & 0 \\[1mm]
-\sqrt{2m\Omega_{12}} & \sqrt{2n\Omega_{34}} & 0 & -g(\sqrt{2}\phi +v) 
\end{array}\right)
\left(\begin{array}{c}\alpha\\ \beta\\ \gamma\\ \delta\end{array}\right)
=\omega \left(\begin{array}{c}\alpha\\ \beta\\ \gamma \\ 
\delta\end{array}\right).
\label{eqb20}
\end{eqnarray}
Using the fact 
\begin{eqnarray}
\det\left(\begin{array}{cccc} K & 0 & A & B \\[1mm]
                              0 & K & B & -A \\[1mm]
                              -A & -B & L & 0 \\[1mm]
                              -B & A & 0 & L
                                \end{array}\right)
=(A^2+B^2+KL)^2,
\label{eqb21}
\end{eqnarray}
we see that the eigenvalue $\omega$ is determined by the following equation,
\begin{eqnarray}
&& \left[\omega +g(\sqrt{2}\phi^*+v)\right]
\left[\omega +g(\sqrt{2}\phi+v)\right]+2m \Omega_{12}
+2n \Omega_{34}=0,\nonumber\\[2mm]
&&\omega_{\pm}(m,n)=-g\left[\frac{\phi+\phi^*}{\sqrt{2}}+v\right]\pm
\sqrt{\frac{1}{2}g^2(\phi-\phi^*)^2-2m \Omega_{12} -2n\Omega_{34}}. 
\end{eqnarray}
Eqs.(\ref{eqb20}) and  (\ref{eqb21}) explicitly show that $\omega_{+}(m,n)$
and  $\omega_{-}(m,n)$ with $m,n{\geq}1$ are doubly degenerate,
since for a $4{\times}4$ matrix there should exist four eigenvalues. Special 
attention should be paid to the cases of $m=0$ or $n=0$ as well as both
of them equal to zero, when we will see that the degeneracies of the eigenvalue 
are different:
\begin{itemize}  
\item $m{\geq}1$, $n=0$:~ in this case the eigenvalue equation (\ref{eqb15})
will reduce to the following form,
\begin{eqnarray}
&&\left(\begin{array}{c} 0 \\[1mm]
\left[-g(\sqrt{2}\phi^*+v)\beta +\sqrt{2m\Omega_{12}}\gamma\right]
|m-1,0{\rangle}\\[1mm]
\left[-\sqrt{2m\Omega_{12}}\beta-g(\sqrt{2}\phi +v)\gamma\right]
|m,0{\rangle}\\[1mm] 0\end{array}\right)
=\omega \left(\begin{array}{c} 0 \\ \beta |m-1,0{\rangle} \\ 
\gamma |m,0{\rangle} \\ 0 \end{array}\right),\nonumber\\
&& \omega_{\pm}(m,0)=-g\left[\frac{\phi+\phi^*}{\sqrt{2}}+v\right]\pm
\sqrt{\frac{1}{2}g^2(\phi-\phi^*)^2-2m \Omega_{12}}.
\end{eqnarray}
The eigenvalues $\omega_{\pm}(m,0)$ are obviously nondegenerate.
\item $m=0$, $n{\geq}1$:~ in this case we have the eigenvalue 
equation as follows,
\begin{eqnarray}
&&\left(\begin{array}{c} 
\left[-g(\sqrt{2}\phi^*+v)\alpha +\sqrt{2n\Omega_{34}}\gamma\right]
|0,n-1{\rangle}\\[1mm] 0\\[1mm]
\left[-\sqrt{2n\Omega_{34}}\alpha-g(\sqrt{2}\phi +v)\gamma\right]
|0,n{\rangle}\\[1mm] 0\end{array}\right)
=\omega \left(\begin{array}{c}  \alpha |0,n-1{\rangle} \\ 0\\ 
\gamma |0,n{\rangle} \\ 0 \end{array}\right),\nonumber\\
&&\omega_{\pm}(0,n)=-g\left[\frac{\phi+\phi^*}{\sqrt{2}}+v\right]\pm
\sqrt{\frac{1}{2}g^2(\phi-\phi^*)^2-2n \Omega_{34}}.
\end{eqnarray} 
The eigenvalues $\omega_{\pm}(0,n)$ are also nondegenerate. 
\item $m=n=0$:~ the eigenvalue equation becomes very simple,
\begin{eqnarray}
\left(\begin{array}{c} 0 \\ 0\\ -g(\sqrt{2}\phi +v) |0,0{\rangle} \\ 0
\end{array}\right)&=&\omega 
\left(\begin{array}{c} 0 \\ 0\\ |0,0{\rangle} \\ 0
\end{array}\right),\nonumber\\
\omega(0,0) = -g(\sqrt{2}\phi +v)=\omega_-(0,0).
\end{eqnarray}
Thus there only exists one $\omega_-(0,0)$ and it is nondegenerate.
\end{itemize}

For the case {\bf 2}, the common eigenstate of $\Delta^+\Delta^-$ and 
$\Delta^-\Delta^+$ with eigenvalue $-2m \Omega_{12}-2n \Omega_{34}$ is
\begin{eqnarray}
\left(\begin{array}{c} |\chi_1{\rangle}\\|\chi_2{\rangle}\end{array}\right)
=\left(\begin{array}{c}\left(\begin{array}{c}
\alpha |m,n{\rangle}\\ \beta|m-1,n-1{\rangle}\end{array}\right)\\
\left(\begin{array}{c}
\gamma |m,n-1{\rangle}\\ \delta|m-1,n{\rangle}\end{array}\right)\end{array}
\right).
\end{eqnarray} 
In a similar way, one can see that the eigenvalues $\omega_{\pm}(m,n)$, 
$\omega_{\pm}(m,0)$, $\omega_{\pm}(0,n)$ with $m,n{\geq}1$ and their 
degeneracies are the same as the case {\bf 1} except that $\omega (0,0)$ is different,
\begin{eqnarray}
\omega (0,0)=-g(\sqrt{2}\phi^* +v)=\omega_+(0,0).
\end{eqnarray} 

As for the cases {\bf 3} and {\bf 4}, the common eigenstates of $\Delta^+\Delta^-$ and 
$\Delta^+\Delta^-$ with eigenvalue $-2m \Omega_{12}-2n \Omega_{34}$ are,
respectively,
\begin{eqnarray}
&3.& ~~
\left(\begin{array}{c} |\chi_1{\rangle}\\|\chi_2{\rangle}\end{array}\right)
=\left(\begin{array}{c}\left(\begin{array}{c}
\alpha |m-1,n-1{\rangle}\\ \beta|m,n{\rangle}\end{array}\right)\\
\left(\begin{array}{c}
\gamma |m-1,n{\rangle}\\ \delta|m,n-1{\rangle}\end{array}\right)\end{array}
\right);\nonumber\\
&4.& ~~ 
\left(\begin{array}{c} |\chi_1{\rangle}\\|\chi_2{\rangle}\end{array}\right)
=\left(\begin{array}{c}\left(\begin{array}{c}
\alpha |m-1,n{\rangle}\\ \beta|m,n-1{\rangle}\end{array}\right)\\
\left(\begin{array}{c}
\gamma |m-1,n-1{\rangle}\\ \delta|m,n{\rangle}\end{array}\right)\end{array}
\right).
\end{eqnarray} 
The eigenvalues $\omega_{\pm}(m,n)$, 
$\omega_{\pm}(m,0)$, $\omega_{\pm}(0,n)$ with $m,n{\geq}1$ and their 
degeneracies are the same as the cases {\bf 1, 2}, but $\omega (0,0)$'s are,
respectively,
\begin{eqnarray}
&3.& ~~\omega (0,0)=-g(\sqrt{2}\phi^* +v)=\omega_+(0,0);\nonumber\\
&4.& ~~\omega (0,0)=-g(\sqrt{2}\phi +v)=\omega_-(0,0).
\end{eqnarray} 

 The eigenvalues of $\widetilde{\Delta}_F$ can be determined in a similar way, and the only difference is $g{\longrightarrow}-g$. 

It should be emphasized that these four cases are not equivalent,
since the eigenstates are different from each other. However, they give  
the identical $\det\Delta_F \det\widetilde{\Delta}_F$.

\end{document}